\documentclass[aps,prl,reprint,floatfix,showpacs]{revtex4-1}

\usepackage{bm}
\usepackage{epsfig}
\usepackage{epstopdf}
\usepackage{graphicx}
\usepackage{graphics}
\DeclareGraphicsExtensions{.eps}
\usepackage{color}

\usepackage{amsmath}
\usepackage{amsfonts}
\usepackage{amssymb}

\begin{document}

\title{Superconducting Gap of UCoGe probed by Thermal Transport}
\author{M.~Taupin$^{1,2}$}%
\author{L.~Howald$^3$}
\email[]{ludovic.howald@physik.uzh.ch}
\author{D.~Aoki$^{1,4}$}
\author{J.-P.~Brison$^1$}
\email[]{jean-pascal.brison@cea.fr}
\affiliation{$^1$Univ. Grenoble Alpes, INAC-SPSMS, F-38000 Grenoble, France\\
$^{\ }$CEA, INAC-SPSMS, F-38000 Grenoble, France\\
$^2$Low Temperature Laboratory, Aalto University, P.O. Box 13500, FI-00076 Aalto, Finland\\
$^3$Swiss Light Source, Paul Scherrer Institut, 5232 Villigen PSI, Switzerland\\
$^4$Institute for Materials Research, Tohoku University, Oarai, Ibaraki 311-1313, Japan}

\date{\today}

\begin{abstract}
Thermal conductivity measurements in the superconducting state of the ferromagnet UCoGe were performed at very low temperatures and under magnetic field on samples of different qualities and with the heat current along the three crystallographic axis. This allows to disentangle intrinsic and extrinsic effects, confirm the situation of multigap superconductivity and shed new light on the situation expected or claimed for the gap in these ferromagnetic superconductors, like evidences of absence of ``partially gapped'' Fermi surfaces.
\end{abstract}


\maketitle

The orthorhombic heavy fermion system UCoGe is one of the few compounds exhibiting long range coexistence between weak itinerant ferromagnetism and superconductivity \cite{Huy2007}. Microscopic coexistence of both orders has been proved by $\mu$SR and NQR measurements on different samples \cite{Visser2009,Ohta2010}, and it is believed that UCoGe, like URhGe and UGe$_2$, is an odd-parity (triplet) $p$-wave superconductor: this claim is supported by the high upper critical field value exceeding the paramagnetic limit \cite{Huy2008,Aoki2009}, and by recent Knight-shift measurements \cite{Hattori2013a}. These results are also in agreement with the more precise prediction, that large band splitting induced by the Fermi surface polarization in the ferromagnetic state, would trigger an ``equal spin pairing (ESP) state'' \cite{Mineev2002, Mineev2004}.
 
Beyond these general features, little is still known on the superconducting order parameter of UCoGe. Early NQR and thermal conductivity measurements suggested possible line nodes of the gap \cite{Ohta2008,Howald2014}, but in samples with large residual terms, suggesting that about $50\%$ of the volume was remaining non superconducting, or dominated by gapless impurity bands. The position of these hypothetical nodes is also undetermined. Owing to the large residual density of states observed in the superconducting state, it has also been claimed that only partial gaping of the Fermi surface is occurring in these ferromagnetic superconductors, and notably in UCoGe \cite{Ohta2010,Aoki2011a}, like in the A1 phase of superfluid $^3$He, or that a self-induced vortex-state is triggered by the ferromagnetic background \cite{Ohta2008,Deguchi2010,Paulsen2012,Aoki2011b,Kusunose2013}. More generally, it has been predicted (but not verified experimentally), that the ESP state should also be a multigap state, with different gap amplitudes on the minority and majority spin bands \cite{Mineev2004}.

We present here results obtained on 5 samples of UCoGe. The choice of this system among the three known ferromagnetic superconductors is motivated by the availability of high quality crystals, together with a bulk superconducting transition temperature $T_{SC}\approx 0.5$\,K at ambient pressure, allowing measurements below $T_{SC}/10$. The measurement of these different crystals allows the distinction of intrinsic and extrinsic behaviors, and the exploration of the anisotropy of the superconducting properties. A main result is the proof of the absence of partially gapped Fermi surfaces in the superconducting state, like the A1 phase of superfluid $^3$He and the confirmation of the recent suggestion reported in \cite{Howald2014}, that UCoGe could have several gaps.%, with the closure of a small gap at approximately 1\,T in the configuration \textbf{H}//\textbf{b}, the second easy-magnetization magnetization axis.

\textit{Experimental}. 
High-quality single crystals were grown using the Czhochralski method in a tetra-arc furnace. All of them are bar shaped, cut along one crystallographic \textbf{a}, \textbf{b} or \textbf{c}-axis, and were characterized by specific heat, Laue X-ray diffraction and resistivity measurements. The effect of the quality on the resistivity and specific heat may be observed in, e.g., references \cite{Aoki2011a,Aoki2014}. The samples are labelled $S^{i}_{x}$, the exponent i is the heat current direction (\textbf{j$_Q$}) and the subscript x is their residual resistivity ratio (RRR, defined as RRR = $\rho$(300K)/$\rho$(T$\to$0K), with $\rho$ the electrical resistivity measured with the current along the \textbf{i}-axis). 
Sample  $S^{c}_{47}$ was cut from a single crystal ingot, after annealing at 1250$^\circ$C for 12h under ultra high vacuum (UHV). The bar shaped sample was subsequently annealed at 900$^\circ$C for 14 days under UHV. The other samples,  $S^{c}_{16}$, $S^{c}_{110}$, $S^{a}_{65}$ and $S^{b}_{150}$ were obtained from other single crystal ingots with annealing at 900$^{\circ}$C for 14 days without high temperature annealing.

Sample $S^{c}_{47}$ appears as the most homogeneous one, even though its RRR is not the highest one. This superior homogeneity is observed both on the sharpness of the ferromagnetic and superconducting resistive transitions and on the specific heat and thermal conductivity anomalies at both transitions. This absence of correlation between improvement of residual resistivity and sharpness of superconducting transitions is well known in heavy fermion systems (an archetypal example is URu$_2$Si$_2$), and is probably related to the fact that residual resistivity is sensitive to ``small size'' defects (impurities, vacancies...) affecting the mean free path, whereas the broadening of transitions is sensitive to large scale defects (like strain induced in the sample, gradients of composition...) inducing a distribution of transition temperatures. %Up to know however, these defects are detected only by low temperature characterization, whereas structural characterization fails to see differences between samples of different qualities. An important aspect of this question is that superconductivity is very sensitive to defects owing to its unconventional nature and to the smallness of the critical temperature: remember that the decrease of the transition temperature $\Delta T_{SC}$ (or equivalently, the appearance of a gapless regime) with the scattering rate $\Gamma$ , as described by the Abrikosov-Gor'kov law, happens in an absolute fashion, not in a relative one ($\Delta T_{SC} \propto \Gamma$, and NOT $\Delta T_{SC}/T_{SC} \propto \Gamma$). 
Even though this sample $S^{c}_{47}$ is labelled as if cut along the \textbf{c}-axis, it should be noted that the high temperature annealing led to a disorientation of the sample, which happens to be cut $30^{\circ}$ off the \textbf{c}-axis in the (\textbf{b},\textbf{c}) plane. It could have happened due to the crossing during the ``high temperature annealing'' of a phase change occurring slightly before the melting temperature \cite{Pasturel}. 

\begin{figure}
		\includegraphics[width=0.48\textwidth]{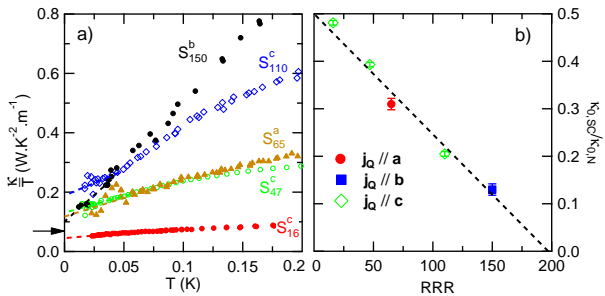}
	\caption{\textbf{a)} Low temperature (below $T_{SC}\approx 0.5$\,K) thermal conductivity of the five samples. Labels indicate heat current direction and RRR values. Note that the data do not converge to a unique (universal) value (see text), indicated by the black arrow, even if restricted to samples cut along the same \textbf{c}-axis. \textbf{b)} Normalized residual thermal conductivity versus RRR, for all 5 crystals, with error bars resulting from the linear extrapolation to $T = 0$\,K of the data of figure \protect{\ref{fig1}}.a).}
	\label{fig1}
\end{figure}

Low temperature thermal conductivity ($\kappa$) measurements were performed down to 10\,mK and up to 8.5\,T with the standard one-heater two-thermometers setup, in the configurations \textbf{H}//\textbf{c}, the easy magnetization axis, and \textbf{H}//\textbf{b}, the second easy magnetization axis. The heat gradient is produced by Joule heating through a 10\,k$\Omega$ resistance and is measured with carbon Matsushita thermometers, which were re-calibrated in situ at each experiment in zero field with a CMN (2Ce(NO$_3$)$_3$-3Mg(NO$_3$)$_2$-24H$_2$O) paramagnetic salt. The results are independent from the applied thermal gradient, which was typically in the range $\Delta T/T\sim 0.1-2\%$. 
% against reference germanium thermometers (T$\ >\ $0.1\,K) and another carbon Matsushita thermometer (T$\ <\ $0.1\,K). The latter is calibrated at each experiment in zero field with a CMN paramagnetic salt. 
 To minimize heat losses, the thermometers and the heater are suspended on the sample holder with Kevlar wires ($20\,\mu$m diameter) and electrical measurements are done with pure NbTi superconducting wires ($35\,\mu$m diameter). The electrical resistivity ($\rho$) was measured with a current of typically 0.1\,mA at high temperature, and was reduced at low temperature to avoid heating due to Joule effect. This Joule effect is mainly dominated by the contact resistances, of order 3 to 10\,m$\Omega$. The Wiedemann-Franz law (WFL) was checked for all measurements in the normal state ($L=\frac{\kappa\rho}{T}$ is equal to the Lorentz number $L_0=2.44 \cdot 10^{-8}$\,W$\Omega$K$^{-2}$ for $T\rightarrow$ 0\,K). Zero field measurements were performed with a zero field cooled magnet to avoid residual magnetic fields. All measurements have been done at fixed field, changed in the normal state to have a more homogeneous field distribution in the sample. As physical properties are strongly angle dependent \cite{Aoki2009}, two goniometers with piezo-electric actuators have been installed, to align precisely and \textit{in-situ} the samples for field measurements along the \textbf{b}-axis.

It has been shown that in UCoGe, magnetic fluctuations are present down to low temperature and contribute to thermal conductivity \cite{Taupin2014}. This was a real surprise, as these contributions exist above $T_{Curie}$, so do not necessarily arise from magnon modes but also from over-damped fluctuations. In any case, these contributions should be removed from the total thermal conductivity to study the superconducting phase: this has been done on all (and only on these) normalized data ($\kappa_{SC}/\kappa_N$) presented here. The magnetic and phonon contributions in the superconducting state have been estimated by extrapolation of the normal state data \cite{Taupin2014}, supposing that the superconductivity does not affect their dispersions. It has been shown however that, except for sample $S^{b}_{150}$, they correspond to less than 15\% of the total thermal conductivity at $T_{SC}$, and that they decrease rapidly on cooling and under field applied along the \textbf{c}-axis \cite{Taupin2014}. Moreover, for all samples, including $S^{b}_{150}$ (with a higher amplitude of the magnetic contributions), the normalized data appear to be weakly dependent of the precise model for the magnetic contribution \cite{Taupin2013}. Normalized data at zero temperature do not depend at all of this extra contribution, which vanishes in this limit: this is expected for a bosonic contribution, and checked experimentally as the Wiedemann-Franz law is verified (above the upper critical field) in UCoGe in the limit of $T\rightarrow 0$\,K \cite{Taupin2014}.

\begin{figure}[ht]
	\centering
		\includegraphics[width=0.45\textwidth]{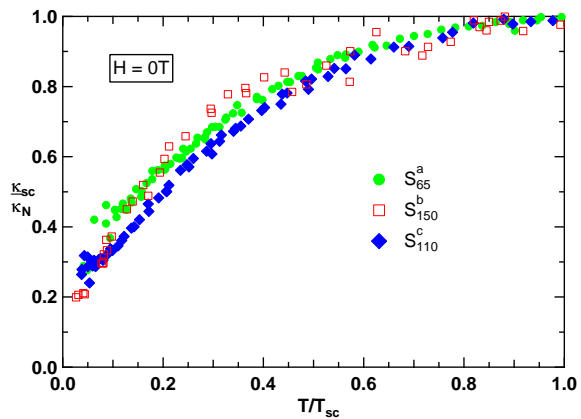}
	\caption{Normalized thermal conductivity versus normalized temperature for the heat current along the three crystallographic directions. The results are similar for the three crystallographic directions, and the difference at the lowest temperature seems to be governed by the residual term, through the RRR (see figure \protect{\ref{fig1}.b)}).}
	\label{fig2}
\end{figure}

\textit{Residual term and sample quality}. 
Figure \ref{fig1}.a) displays the thermal conductivity divided by the temperature ($\kappa/T$) at zero field of the five samples up to 0.2\,K (a wider temperature range of the measurements has already been presented in \cite{Taupin2014}). 
%The superconducting transition manifests itself by a kink in sample $S^{c}_{16}$ and by a broad maximum in samples $S^{a}_{65}$, $S^{c}_{110}$ and $S^{b}_{150}$, sample $S^{c}_{47}$ showing a sharp maximum. The different behaviors in the normal state between the samples arise from the different balance between elastic and inelastic scattering among the samples (see \cite{Taupin2014} for quantitative analysis).
Note that the residual thermal conductivity changes with the RRR. Particularly, a linear extrapolation of the thermal conductivity gives $\frac{\kappa}{T}\big|_{T\to\,0K}$ = 0.19, 0.13 and 0.046\,WK$^{-2}$m$^{-1}$ for samples $S^{c}_{110}$, $S^{c}_{47}$ and $S^{c}_{16}$ respectively, i.e. for the given heat current direction \textbf{j$_Q$}//\textbf{c}. This indicates that the regime of the universal limit, expected for \textit{d}-wave superconductors when the heat current is in the nodal direction, is not reached in our measurements \cite{Graf1996,Shakeripour2009}. A rough estimation of the expected value (from the formulas in \cite{Graf1996}), shown by an arrow on the figure \ref{fig1}.a), is around 0.07\,WK$^{-2}$m$^{-1}$, if the relative slope of the gap at the putative nodes \cite{Graf1996} is equal to 1. This violation of the ``universal limit'' may be due to the fact that the samples are still too inhomogeneous (normal parts remain in the sample), or more likely, that the universal limit does not exist in this compound (if for example, the gap vanishes at point nodes along the \textbf{c}-axis, or if the compound is multigaped).

It should be noticed that the residual term, in ferromagnetic superconductors, magnetic domain walls and self induced vortices can be present even in zero field. For a superconductor in the clean limit, vortex core contribution is known to be negligible, and the main effect should be on the scattering mechanism. But even in the best samples, the mean free path remains governed by impurity (or defect) scattering the internal field is at most 10\,mT \cite{Paulsen2012}, so we do not expect to observe any measurable effect on the thermal transport.

The normalized residual thermal conductivity is defined as $\frac{\kappa_{0,SC}}{\kappa_{0,N}}=\ $lim$_{T\to0\,K}\frac{\kappa_{SC}}{\kappa_{0,N}}$ and $\frac{\kappa_{0,N}}{T}=\frac{L_0}{\rho_0}$, $\rho_0$ the residual electrical resistivity. It is plotted versus the RRR in figure \ref{fig1}.b): it decreases linearly with the RRR, highlighting that the superconducting ground state is strongly dependent on the density of impurities and defaults, as expected for unconventional superconductivity (see e.g. \cite{Balatsky2006}). The decrease of $\kappa_{0,SC}/\kappa_{0,N}$ also highlights the fact that no Fermi surface remains intrinsically normal in the superconducting state. This contradicts theories claiming that superconductivity in UCoGe may be described like the A1 phase of superfluid $^3$He \cite{Tada2011}, based on the results of the early NQR experiments \cite{Ohta2010,Ohta2008}. Indeed, as the proportion of minority spins compared to the majority spins should not depend on the RRR, the normalized residual thermal conductivity, if it was controlled by an un-gapped Fermi surface of one of the spin species, should also be RRR independent.

\begin{figure}
	\centering
		\includegraphics[width=0.48\textwidth]{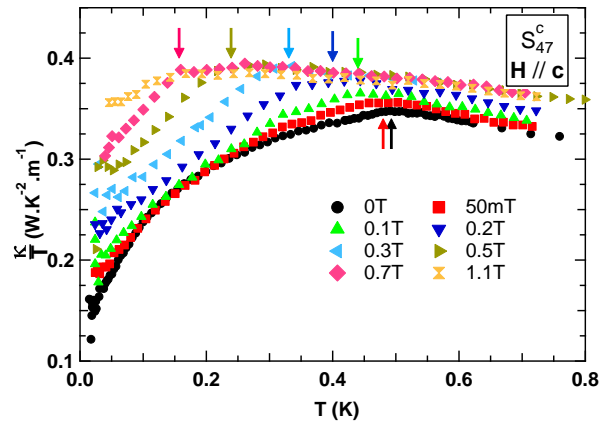}
	\caption{Temperature dependence of the thermal conductivity of sample $S^{c}_{47}$ at several fields. The arrows correspond to the superconducting temperature. Note the change of regime below 0.1\,K in zero field, which is quickly suppressed by fields much lower than $H_{c2}$.}
	\label{fig3}
\end{figure}

\textit{Gap anisotropy}.
The normalized thermal conductivity in the superconducting state is displayed in figure \ref{fig2}, for the heat current applied in the three crystallographic directions (samples $S^{a}_{65}$, $S^{b}_{150}$ and $S^{c}_{110}$). The normalized thermal conductivity in the superconducting state is quite similar for the three heat current directions: the residual term seems directly proportional to the RRR of the sample, whatever the current direction (see figure \ref{fig1}.b)), and a linear increase of $\kappa_{SC}/\kappa_{N}$ with temperature for $T \lesssim$ 0.4 $T_{SC}$. Such a temperature dependence is often claimed to be an (indirect) proof of line of nodes in the current direction, due to the effects of impurities for such a gap topology. However, it is always difficult, if not impossible, to distinguish between a residual term due to (homogeneous) gapless behavior arising from impurity bands \cite{Lee1993}, or from a distribution of critical temperatures in the sample including non superconducting regions: for example, such a linear behavior has been measured in the skuterrudite PrOs$_4$Sb$_{12}$ \cite{Seyfarth2005}, until sample improvement revealed a nodeless gap with a $T^3$ (phonon dominated) temperature dependence of the heat transport.  In UPt$_3$, the first thermal conductivity measurements gave an isotropic and linear thermal conductivity ($\kappa/T$) in the superconducting state \cite{Behnia1991} and much better quality samples were required to observe the ``intrinsic'' $T^2$-behavior \cite{Lussier1996, Suderow1997} arising from line nodes. At the present stage, our measurements do not reveal a pronounced nodal anisotropy of the superconducting gap, despite  theoretical predictions for orthorhombic $p$-wave superconductors suggesting a very anisotropic nodal structure, or no nodes at all \cite{Mineev2002, Mineev2004}.

\textit{Thermal spectroscopy}. 
Sample $S^{c}_{47}$ was measured under field applied along the \textbf{c}-axis. The relatively low value of the upper critical field ($H^{c}_{c2}\approx1$\,T) along this direction allows us to probe the entire superconducting phase. Such a study has been performed on this sample as it is the most homogeneous one. The temperature dependence at a given field is shown on figure \ref{fig3}. The measurement at 1.1\,T is entirely in the normal phase. The arrows point to the superconducting temperature at each field. The superconducting transition has been determined as the onset of the deviation of the Lorenz number ($L=\frac{\kappa\rho_N}{T}$ and $\rho_N=\rho_0+AT^2$) from its (linear) extrapolation to $L_0$ for $T\rightarrow 0$\,K. In the superconducting state, the zero field curve seems to display two regimes: the thermal conductivity decreases monotonously in the whole temperature range below $T_{SC}$, but the decrease is more pronounced and linear below 0.1\,K. This change of regime disappears very rapidly for small applied fields: for fields higher than 0.2\,T, the thermal conductivity decreases linearly from $T_{SC}$ down to the lowest measured temperatures. 

\begin{figure}
	\centering
		\includegraphics[width=0.48\textwidth]{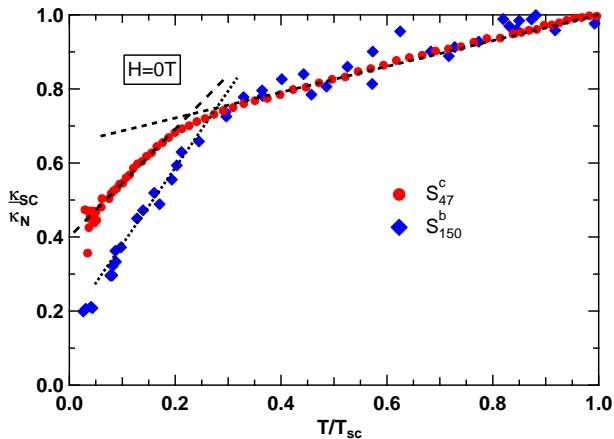}
	\caption{Temperature dependence of the normalized thermal conductivity versus the normalized temperature  for samples $S^{c}_{47}$ and $S^{b}_{150}$ at zero field. The change of regime at $\approx\ 0.2\,T_{SC}$, already apparent on figure \protect{\ref{fig3}}, shows up even more clearly after suppression of the magnetic and phonon contribution, and with the normalization to the normal state extrapolated value. It is a strong support for multigap superconductivity. Dashed lines are guides to the eyes.}
	\label{fig4}
\end{figure}

Interestingly, the change of regime in zero field is even more striking on the normalized thermal conductivity data, which is displayed on figure \ref{fig4}. In this latest figure, we display also the same normalized data for sample $S^{b}_{150}$, cut along the \textbf{b}-axis, which has the smallest residual term of all measured samples (same data as on figure \ref{fig2}): clearly, an even stronger decrease of $\kappa_{SC}/\kappa_{N}$ is observed below 0.2\,$T_{SC}$ on this sample, due to its smaller residual term. Although less visible, when homogeneity or purity are weaker, this change of regime at 0.2\,$T_{SC}$ is also detected in the other samples. Both this change of regime and its disappearance for fields much lower than $H_{c2}$ are naturally explained by multigap superconductivity (see for example the specific heat data in MgB$_2$ \cite{Bouquet2001}), and could only be explained by an additional low temperature phase transition in single gap superconductors. For example, the system could go from a A1-like superconducting phase (electrons partially condensed) to a fully condensed state below a certain temperature. This is however unlikely as, in superfluid  $^3$He, a specific heat anomaly has been observed at the transition from the A1 to the A2 phase \cite{Halperin1976}. By analogy, a thermodynamic signature would be expected in UCoGe at $0.2\,T_{SC}$.

\begin{figure}
	\centering
		\includegraphics[width=0.48\textwidth]{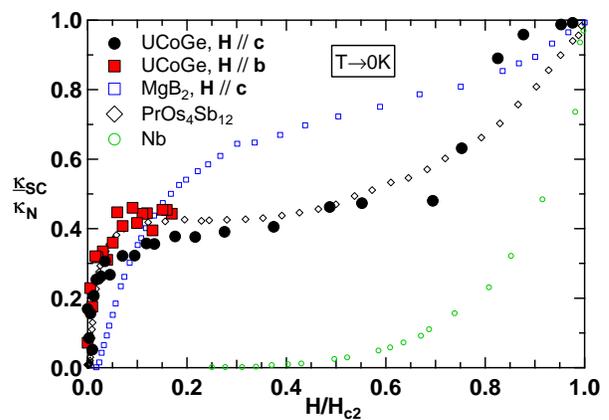}
	\caption{Full symbols: the normalized thermal conductivity versus normalized field extrapolated at zero temperature when the field is applied along the \textbf{b} and the \textbf{c}-axis of sample $S^{b}_{150}$. The normalizing (upper critical) field along the \textbf{c}-axis is $H^c_{c2}=1$\,T and, along the \textbf{b}-axis, has been extrapolated to $H^b_{c2}=50$\,T (see text for explanation). The figure displays also the thermal conductivity of Nb \protect{\cite{Lowell1970}}, PrOs$_4$Sb$_{12}$ \protect{\cite{Seyfarth2006}} and MgB$_2$ \protect{\cite{Sologubenko2002}}%, with \textbf{H}//\textbf{c}
	.}
	\label{fig5}
\end{figure}

\textit{Field spectroscopy}. 
Measurements up to 8.5\,T with the field along the \textbf{b} and the \textbf{c}-axis were performed on sample $S^{b}_{150}$. The normalized thermal conductivity is plotted against the normalized field on figure \ref{fig5} in the limit $T\to$ 0\,K. The critical fields for normalization have been taken as $H^c_{c2}=1$\,T for the \textbf{c}-axis and $H^b_{c2}=50$\,T for the \textbf{b}-axis. This value of $H^b_{c2}$ is not the true upper critical field at zero temperature, but the extrapolation of the upper critical field from its initial slope at $T_{SC}$ ($H$ $<$ 8\,T), which is more relevant for this purpose: at higher field values, $H^b_{c2}$ displays an ``S''-shape \cite{Aoki2009}, indicating an unusual change of the pairing strength or of the normal state under field (like that expected for a topological change of the Fermi surface due to a Lifshitz transition \cite{Malone2012}). %So the low field ($H$ $<$ 8\,T) regime should not be renormalized by values deduced from the high field regime.

It appears clearly on figure \ref{fig5} that for both field configurations, two energy scales are found: first a rapid increase from 0 to $H/H_{c2}\approx 0.05$ followed by a plateau at the values $\kappa_{SC}/\kappa_N\approx0.35$ in the configuration \textbf{H}//\textbf{c}, and $\kappa_{SC}/\kappa_N\approx0.45$ when \textbf{H}//\textbf{b}. A second step is found at $H/H_{c2}=0.8$ when \textbf{H}//\textbf{c}. By comparison, the thermal conductivity of the one band superconductor Nb \cite{Lowell1970} and of the two band superconductors MgB$_2$ \cite{Sologubenko2002} and PrOs$_4$Sb$_{12}$ \cite{Seyfarth2006} are displayed on the same graph. The two-step behavior of the field dependence of the zero temperature $\kappa_{SC}/\kappa_N$ is typical of multigap superconductors. The first step arises from the suppression of the small gap by an effective upper critical field much smaller than the true $H_{c2}$. Let us note that a Doppler shift of the excitation spectrum of a single band superconductor would be governed only by a single energy scale ($H_{c2}(0)$), and would never explain a recovery of $40\%$ of the normal state value on such a small renormalized field range. So both the very strong increase at low field, and the plateau, pointing to a second field scale, are a hallmark of multigap superconductivity. Such a trend was already observed in sample $S^c_{16}$ \cite{Howald2014}, but only for the configuration \textbf{H}//\textbf{b}, probably due to the too large residual term compared to the contribution of the small gap band. Moreover like in PrOs$_4$Sb$_{12}$ \cite{Seyfarth2006}, the consistency between the thermal (figure \ref{fig4}) and magnetic field (figure \ref{fig5}) behavior of $\kappa/T$ is an additional strong support for the conclusion that UCoGe is indeed a multigap superconductor.

\textit{Conclusion}.
This study of crystals of different quality and orientations could firmly establish that UCoGe is certainly not in the same regime as the A1 phase of superfluid $^3$He, but is clearly a multigap superconductor. This is also a good reason for not reaching the universal conductivity regime (as demonstrated for transport along the \textbf{c}-axis). The origin of this multigap superconductivity is however elusive: is it controlled by a change of coupling strength owing to the $f$-character of the bands \cite{Seyfarth2006, Seyfarth2008}, or by the spin state (majority/minority) of the underlying ferromagnetic ground state? The very strong upward curvature of the upper critical field along the \textbf{c}-axis, which is very unusual \cite{Aoki2009}, should be revisited in the light of this two band model, and might give a clue to this question. A second open question is to determine the symmetry of the superconducting order parameter in this ferromagnetic superconductor, and notably to probe if theoretical predictions of a very anisotropic nodal structure (if any) are verified: up to now, a rather isotropic behavior of $\kappa/T$ has been found, pointing either to an isotropic nodal structure, or more likely to dominant (isotropic) impurity effects. The solution to this problem is certainly lying in crystals of further improved RRR, which might be at reached for this system.
%: the race is now open to reach RRR values above 200, which could be enough to observe the intrinsic (low) temperature dependence of $\kappa/T$, and its anisotropy. 

\section{Acknowledgement}
We acknowledge very fruitful discussions with V. Mineev, G. Knebel and J. Flouquet. This work was supported by the French ANR grant SINUS and the ERC grant ``NewHeavyFermion''.

%\begin{thebibliography}{99}

%merlin.mbs apsrev4-1.bst 2010-07-25 4.21a (PWD, AO, DPC) hacked
%Control: key (0)
%Control: author (72) initials jnrlst
%Control: editor formatted (1) identically to author
%Control: production of article title (-1) disabled
%Control: page (0) single
%Control: year (1) truncated
%Control: production of eprint (0) enabled
%

%\bibliography{biblio,/Users/jeanpascalbrison/biblioLatex/bibliography}
%\bibliography{biblio}

%\bibitem{PasturelPrivate} Mathieu Pasturel, Sciences Chimiques de Rennes, France, private communication.

%\end{thebibliography}
\end{document}